\documentclass[%
nofootinbib,
superscriptaddress,
 amsmath,amssymb,
 aps,
pra,
 longbibliography,
 floatfix,
 lengthcheck,%
]{revtex4-1}

\usepackage{hyperref}
\usepackage{RabitzGate_SCI_T}

\newcommand{\Tr}{\mathrm{Tr}}

\hypersetup{
  colorlinks=true,
  linkcolor=blue!50!red,
  urlcolor=red!70!black
}

\usepackage{soul}
\newcommand{\citeurl}[1]{See {\href{#1}{\expandafter\string #1}}.}

\begin{document}

\title[Comment on ``Control landscapes are almost always trap free...'']{Comment on ``Control landscapes are almost always trap free: a geometric assessment''}

\author{Dmitry V. Zhdanov}
\address{University of Bristol, Bristol BS8 1QU, UK}
\email{dm.zhdanov@gmail.com}

\begin{abstract}
We analyze a recent claim that almost all closed, finite dimensional quantum
systems have trap-free (i.e., free from local optima) landscapes (B. Russell \emph{et al} 2017 \emph{\href{https://doi.org/10.1088/1751-8121/aa6b77}{\textnormal{ J. Phys. A: Math. Theor. \textbf{50}, 205302}}}). We point out several errors in the proof which compromise the authors' conclusion.
\end{abstract}

\maketitle
\onecolumngrid
\vspace{-0.2pc}{\color{blue}{\noindent \it Update:} Interested readers are highly encouraged to take a look at the ``rebuttal'' \cite{2018-Russell-a} of this comment published by the authors of Ref.~\cite{2017-Russell}. This ``rebuttal'' is a showcase of the way the erroneous and misleading statements under discussion will be wrapped up and injected in their future works, such as Ref.~\cite{2018-Kosut}.
}
\vspace{0.5pc}

\vspace{0.5pc}
\twocolumngrid
%
%
%
%

\section{Introduction}
Finding ``fast'' algorithms capable of resolving complicated combinatorial problems in reasonable time is one of the primary challenges of optimal control theory. It was found that a number of ``hard'' optimization problems carry the following fortunate phase transition property: When the ratio $\beta{=}\frac KN$ between the numbers of control parameters $K$ and constraints $N$ exceeds a certain threshold $\beta_{\idx{tr}}$, then there exist natural gradient algorithms resolving the optimization problem within the time polynomial in $N$ \cite{2014-Vakulenko}. A conceptual possibility of such phase transition in the quantum optimal control of closed systems (QOC-CS) follows from classical works \cite{1937-von_Neumann,1988-Brockett}. However, no explicit generic theoretical evidences were reported until the recent paper \cite{2017-Russell} by Russel et.al. The authors of Ref.~\cite{2017-Russell} made the following rather strong claim:
\begin{doubted_proposition}\label{@P:key_result}
Phase transition with very small threshold ($\beta_{\idx{tr}}{\ll}1$)  is a generic feature of nearly every QOC-CS problem.
\end{doubted_proposition}
Since classical mechanics is merely a limiting case of quantum mechanics, the reported result has pivotal practical implications. Namely, the authors argue in their subsequent publications \cite{2017-Russell2,2018-Russell} that optimal control problems in nearly all areas, from chemistry and material science to biological evolution, fundamentally belong to a ``simple'' category and can be solved in polynomial times using a relatively small number of controls.

Below we will analyze in detail the proof of proposition \ref{@P:key_result} and will show that it is incorrect. The presentation is organized as follows. In the next section~\ref{@SEC:background} the basics of QOC-CS are outlined to the extent necessary to formulate the central result of work~\cite{2017-Russell}, namely, theorem 4.2. This theorem with clarifying comments is then presented in Sec.~\ref{@SEC:theorem_4.2}. The subsequent section~\ref{@SEC:criticism} is the core of the present comment. There, we identify two mistakes in the proof of theorem 4.2 and provide counterexamples showing that its statement is incorrect. The paper concludes with an outlook of recent literature results derived from theorem 4.2 which require revision in the light of this comment.

To avoid confusions, we will use flags \wrong and \doubted to explicitly distinguish the statements from work \cite{2017-Russell} in which proofs we have identified a mistake. The flag \wrong will mark statements falsified by explicitly constructed counterexamples. The flag \doubted will mark still possibly correct statements which can be regarded as a conjectures.

All other justifiably valid statements will be marked using \correct flag.

\section{Problem settling and necessary definitions\label{@SEC:background}}

Consider the following terminal optimal control problem for a closed quantum system in $N$-dimensional Hilbert space:
\begin{align}\label{__Objective}
J{=}\midop{O}{=}\Tr[{\hat O}{\hat\rho(T)}]{\to}\max_{\eee{\in}{\manE}}.
\end{align}
Here $J$ is the control objective, $\hat O$ is some quantum-mechanical observable and $\eee$ is the set of time-dependent control parameters which guide evolution of system density matrix $\hat\rho(t)$ from given initial state $\hat\rho(0){=}\hat\rho_0$ at the time $t{=}0$ to the final state $\hat\rho(T)$ at $t{=}T$: 
\begin{align}\label{__Evolution}
\hat\rho(T){=}\hat U_T(\eee)\hat\rho_0\hat U_T^{\dagger}(\eee).
\end{align}
The unitary operator $\hat U_t(\eee)$ in equation~\eqref{__Evolution} satisfies the evolution equation
\begin{gather}
\der{}{t}U_t(\eee){=}{-}\frac{i}{\hbar}\hat H(\eee,t)U_t(\eee)
\end{gather}
and the initial condition $\hat U_0(\eee){=}\hat I$, where $\hat I$ is identity operator and $\hat H(\eee,t)$ is the controlled  system Hamiltonian.

In order to proceed, few definitions are needed. 

Given two smooth manifolds $\manA$ and $\manB$, the map $f:\manA{\to}\manB$ is called \emph{globally surjective} if for each point $b{\in}\manB$ there exist at least one point $a{\in}\manA$ such that $f(a){=}b$. If each point $b'{\in}\manB$ from a tiny neighborhood of $b$ has a preimage $a'$ in a tiny neighborhood of $a$ then the map $f$ is called \emph{locally surjective} at $a$. 

The set of control-dependent unitary operators $\hat U_T(\eee)$ defines the map 
\begin{gather}\label{__psi-map}
\psi: \manE {\to} SU(N)
\end{gather} between space of controls and special unitary matrices. A system is called \emph{controllable} if the map $\psi$ is globally surjective.\footnote{In other words, the system is globally controllable if for any unitary operator $\hat U'$ there exist at least one set of controls $\eee'$ such that $\hat U_T(\eee'){=}\hat U'$} A system is called locally controllable if the map $\psi$ is also everywhere locally surjective.

The function $J(\eee)$ defined by equations \eqref{__Objective} and \eqref{__Evolution} is sometimes called \emph{quantum control landscape}. It maps the space of controls $\manE$ to range $\manO$ of admissible values of $\midop{\hat O}$
\footnote{The manifold $\manO$ is an interval $[\sum_{i{=}1}^Nr_{N+1-i}o_i,\sum_{i{=}1}^Nr_io_i]$, where $o_i$ and $r_i$ are the eigenvalues of $\hat O$ an $\hat\rho_0$ enumerated in increasing order \cite{1937-von_Neumann}.}%
. 
If the map $J:\manE{\to}\manO$ is locally surjective then one is guaranteed to reach the global maximum of $J(\eee)$ by iterative small variations of control parameters using a gradient algorithm. In other words, local surjectivity implies above-threshold case $\beta{>}\beta_{\idx{tr}}$.

The following fundamental result is proven in classical works \cite{1937-von_Neumann,1988-Brockett}.
\begin{correct_theorem}\label{__theo_Brockett}
The map $\phi_{\hat O}:SU(N){\to}\manO$ defined as $\phi_{\hat O}(\hat U){=}\Tr[\hat O\hat U\hat\rho_0\hat U^{\dagger}]$ is localy surjective for any $\hat O$ and $\hat\rho_0$.
\end{correct_theorem}
\begin{correct_corollary}\label{__cor_Brockett}
Local surjectivity of map \eqref{__psi-map} is a sufficient condition for local surjectivity of quantum control landscape $J(\eee)$.
\end{correct_corollary}

\section{Rigorous formulation of the claim of paper \texorpdfstring{\cite{2017-Russell}}{}
\label{@SEC:theorem_4.2}}

The central result of paper \cite{2017-Russell} is theorem 4.2. Both its formulation and proof rely on important premise made in the beginning of section 4.1 which in fact constitutes the following additional theorem:  
\begin{doubted_theorem}\label{__theo_rsl_a}
For any $N$-dimensional closed quantum system and substantially large natural number $Z$ it is possible to formally introduce a set $\eee$ of bounded controls $\ee_{j}{\in}[-\kappa,\kappa]$ $(j{=}1,...,(N^2{-}1)Z)$, such that the system is controllable and function $J(\eee)$ is locally surjective.
\end{doubted_theorem}
Specifically, the authors of Ref.~\cite{2017-Russell} claim in the paragraph following equation~(12) that the set $C_{\kappa,p}$ of $\kappa$-bounded piecewise-constant controls introduced by equation~(10) of their work satisfies the statement of above theorem (and hence proves it). The critical part of this claim can be compactly reformulated as the following lemma.
\begin{wrong_lemma}\label{__theo_rsl_a'}
Consider the controlled Hamiltonian
\begin{gather}\label{__ham_rsl}
\hat H(\eee,t){=}
\sum_{j{=}1}^{N^2-1}\sum_{z{=}1}^{Z}\ee_{j,z}b_z(t)\hat B_j,
\end{gather}
where the boxcar functions $b_k(t)$ are nonzero (equal to 1) only on $z$-th time subinterval $(\frac{z-1}ZT,\frac{z}ZT]$ and $\ee_{j,z}{\in}[-\kappa,\kappa]$ represent the set $\eee$ of $(N^2{-}1)Z$ bounded control parameters. The $j$-summation runs over a complete basis $\{\hat B_j\}$ of $SU(N)$.

Assume that 
\begin{enumerate}
\item the duration $\frac{T}Z$ of each piecewise-constant segment of the control is short enough and the speed of the curve $\hat U_t(\eee)$ is not too high (i.e., $\kappa$ is small enough), so that the following inequality is satisfied:
\begin{gather}\label{__kappa_constraint}
\frac{T}Z{<}\frac{2\pi}{E_{\idx{max}}(\kappa){-}E_{\idx{min}}(\kappa)},
\end{gather}
where $E_{\idx{max}}(\kappa)$ and $E_{\idx{min}}(\kappa)$ are maximal and minimal eigenvalues of $\hat H(\eee,t)$ over an admissible control domain $\eee{\in}\manE(\kappa)$;
\item $\kappa$ is large enough, so that the map $\psi$ defined by equation~\eqref{__psi-map} is globally surjective.
\end{enumerate}
Then, the map $\psi$ is also locally surjective.
\end{wrong_lemma}
The proof of theorem~\ref{__theo_rsl_a} readily follows from combining the above lemma with corollary~\ref{__cor_Brockett}.

The next theorem reproduces the remaining relevant part of theorem 4.2.
\begin{wrong_theorem}\label{__theo_rsl_b}
For a controllable system which landscape $J(\eee)$ is locally surjective everywhere in the control space $\manE$, fixing any single control parameter $\ee_{j'}{=}c$ may introduce local maxima and minima into the new control landscape $J(\eee|\ee_{j'}{=}c)$ (a function of the remaining unfixed variables $\ee_{j{\ne}j'}$ only) only for a null set of values of $c{\in}\mathbb{R}$.
\end{wrong_theorem}

It is noteworthy that theorem~\ref{__theo_rsl_b} may be applied to any controllable system with locally surjective landscape $J(\eee)$. Its special relevance for quantum control problems leans upon theorem~\ref{__theo_rsl_a} which claims that the required local surjectivity of $J(\eee)$ is a generic feature for this class of problems. 

Proposition \ref{@P:key_result} can be derived from theorems \ref{__theo_rsl_a} and \ref{__theo_rsl_b} as follows. First, one introduces a very rich set of controls $\eee$ satisfying theorem \ref{__theo_rsl_a}. Lemma \ref{__theo_rsl_a'} guarantees that it is always possible. Second, one starts to ``freeze'' controls $\ee_j$ one by one. Theorem \ref{__theo_rsl_b} implies that such control elimination is highly unlikely to break the local surjectivity of control landscape. Hence, the iterative control eliminations can be repeated until they will start compromising the system controllability. Thus, we can ``freeze'' most of controls $\ee_{j}$ at arbitrary values without destroying the local surjectivity of control landscape. The latter implies that the system remains optimizable by gradient methods. Since the number of remaining controls is expected to be small we can conclude that $\beta_{\idx{tr}}{\ll}1$ which justifies proposition \ref{@P:key_result}.

\section{Criticism\label{@SEC:criticism}}
In this section, we are going to show that the proofs of both theorems \ref{__theo_rsl_a} and  \ref{__theo_rsl_b} contain mistakes. As a result, the validity of theorem~\ref{__theo_rsl_a} becomes an open question. However, the incorrectness of theorem \ref{__theo_rsl_b} will be justified by explicit counterexamples.

\subsection{Theorem \ref{__theo_rsl_a}}
The validity of theorem \ref{__theo_rsl_a} is compromised by incorrect lemma \ref{__theo_rsl_a'}. A weak point in its proof is that variations of controls at the boundary of $\manE$ are more constrained compared to its interior. These additional boundary constraints can break local surjectivity of the map $\psi$, as will be shown by example. 

For simplicity, we will consider two-level system for which the basis operators $\hat B_j$ in equation \eqref{__ham_rsl} are represented by Pauli matrices $\hat\sigma_i$. Hereafter we will set $\hbar{=}1$. Denote as $\kappa_{\idx{ctr}}(T,Z)$ the minimal value of boundary constraint $\kappa$, such that the system is globally controllable (i.e., the map $\psi$ is globally surjective) for any $\kappa{>}\kappa_{\idx{ctr}}(T,Z)$. Let us also introduce the threshold parameter $\kappa_{\idx{thr}}(T,Z)$, such that inequality \eqref{__kappa_constraint} is satisfied for $\kappa{<}\kappa_{\idx{thr}}(T,Z)$. We will choose $Z$ large enough, so that $\kappa_{\idx{thr}}(T,Z){\gg}\kappa_{\idx{ctr}}(T,Z)$ and consider the case
$\kappa_{\idx{ctr}}(T,Z){<}\kappa{<}\kappa_{\idx{thr}}(T,Z)$. This way, we ensure that our system satisfies all the conditions of lemma \ref{__theo_rsl_a'}.

Denote as $\eee'$ the control policy where all the controls are set to their maximal values: $\ee_{j,z}{=}\kappa$. The corresponding Hamiltonian \eqref{__ham_rsl} takes the time-independent form
\begin{gather}
\hat H(\eee',t){=}\sum_{j{=}1}^{3}\kappa\hat \sigma_j.
\end{gather}

It is straightforward to check (using, e.g., Pontryagin maximum principle) that the controls $\eee'$ correspond to a local maximum (or ``trap'') in the optimization problem \eqref{__Objective} with $\hat O{=}\sin(\alpha{+}\frac{\pi}{3})\hat\sigma_1{+}\sin(\alpha{-}\frac{\pi}{3})\hat\sigma_2{+}\sin(\alpha)\hat\sigma_3$ and $\hat\rho(0){=}\frac12(\hat I{+}\hat\sigma_3)$, where $\alpha{=}2\sqrt{3}T\kappa$. This is a clear sign that the map \eqref{__psi-map} is not locally surjective at $\eee'$. For example, if $\alpha (\mathrm{mod}~2\pi){=}0$, so that 
$\hat U_T(\eee'){=}(-1)^{\frac{\alpha}{2\pi}}\hat I$,
then no infinitesimal variation of controls $\delta\eee$ allows to obtain any of the unitary operators
\begin{gather}
(-1)^{\frac{\alpha}{2\pi}}e^{\epsilon_1\hat\sigma_1{+}\epsilon_2\hat\sigma_2}.
\end{gather}
with infinitesimal $\epsilon_1{>}0$ and $\epsilon_2{>}0$.

Breakdown of the local surjectivity of map $\psi$ at the boundary of $\manE$ is very generic property. However, we hypothesize that the statement of lemma~\ref{__theo_rsl_a'} can be a workable \emph{approximation} when the basins of convergence to the associated boundary traps in the gradient search are small. The latter is likely to be the case when $\kappa{\gg}1$ and $Z{\gg}1$. However, further research is needed to judge the validity of this hypothesis and to identify its scope of applicability.

\subsection{Theorem \ref{__theo_rsl_b}}

The proof of theorem \ref{__theo_rsl_b} proposed in Ref.~\cite{2017-Russell} is based on certain results in differential geometry. For this reason, it is worth to recall a few standard definitions and theorems from this field.

Two submanifolds $\manA$ and $\manB$ of a given finite-dimensional smooth manifold $\manC$ are said to intersect \emph{transversally} if either 1) they do not intersect, or 2) at every point $p$ of intersection, their separate tangent spaces at that point together generate the tangent space of the ambient manifold $\manC$ at that point:
\begin{gather}
T_p(\manA) \oplus T_p(\manB) = T_p(\manC).
\end{gather}

Let $f: \manA{\to}\manB$ be a smooth map of a manifold $\manA$ to a manifold $\manB$, 
containing a smooth submanifold $\manC$. The map $f$ is said to be \emph{transversal} to $\manC$ 
at the point $a$ of $\manA$ if either $f(a)$ does not belong to $\manC$ or the image of 
the tangent space to $\manA$ at $a$ under the derivative $f_{*a}$ is transversal to the tangent space to $\manC$:
\begin{gather}
f_{*a}T_{a}(\manA) \oplus T_{f(a)}\manC = T_{f(a)}{\manB}.
\end{gather}


\begin{correct_theorem}(Parametric transversality theorem)\label{__theo_PT}
Consider a smooth map $F:\manA{\times}\manEe{\to}\manB$ of the direct product of smooth manifolds $\manA$ and $\manEe$ to a smooth manifold $\manB$. We shall consider $F$ as a family of maps $F_{e}$, of manifold $\manA$ to $\manB$, depending on the point $e$ of the manifold $\manEe$ as on a parameter. Then, if the map $F$ is transversal to the submanifold $\manC$ of the manifold $\manB$, almost every member $F_{e}:\manA{\to}\manB$ of the family is transversal to $\manC$.
\end{correct_theorem}

The authors of Ref.~\cite{2017-Russell} apply theorem \ref{__theo_PT} as follows. First, they represent manifold of control parameters as a Cartesian product $\manE{=}\manE_{j}{\times}\manE_{-j}$, where $\manE_{j}$ defines the domain of $j$-th control parameter only and $\manE_{-j}$ includes the possible combinations of all control parameters except $j$-th. Next, for each $J_0{\in}\manO$ they define the manifold $\manL_{J_0}{\subset}SU(N)$ as a subset of all unitary matrices $\hat U{\subset}SU(N)$, such that $\phi_{\hat O}(\hat U){=}J_0$. Finally, they apply theorem \ref{__theo_PT} with $F{=}\psi$, $\manA{=}\manE_{-j}$, $\manEe{=}\manE_{j}$, $\manB{=}SU(N)$ and $\manC{=}\manL_{J_0}$. The proper conclusion should be that the constrained map $\hat U_T(\eee|\ee_j{=}c)$ is transversal to any submanifold $\manL_{J_0}$ for almost all values of $c$. Using corollary \ref{__cor_Brockett}, this result can be equivalently re-expressed in more physical terms as the following theorem.
\begin{correct_theorem}
\label{__theo_PT2}
Let $J(\eee)$ be a landscape having no local extrema. Then, for any given control index $j$ and real number $J_0$ there may be only a null set $\manC$ of values $c$ for which the constrained landscape $J(\eee|\ee_j{=}c)$ includes such points $\eee'$ that: 1) $\eee'$ is a local extremum of $J(\eee|\ee_j{=}c)$, and 2) $J(\eee'|\ee_j{=}c){=}J_0$.
\end{correct_theorem}

The authors derive theorem~\ref{__theo_rsl_b} by claiming that theorem \ref{__theo_PT2} implies that the constrained landscape $J(\eee|\ee_j{=}\cc)$ has no local extrema for almost all values of $c$. However, the claim is incorrect. Theorem~\ref{__theo_PT2} merely implies that the subset of critical values $J_0{\in}\manO$ for the constrained landscape $J(\eee|\ee_j{=}\cc)$ is a null set in $\manO$ (cf. Bertini-Sard theorem \cite{1985-Arnol'd}, p.~32). Informally, this means that the points which are either local or global extrema of the constrained landscape are much less abundant than all remaining points. However, this is very generic property of any smooth function. For instance, the subset of critical values for function $J(\ee){=}\sin(\ee)$ is a null set consisting of just two points $J_0{=}{\pm}1$ despite this function has infinite number of minima and maxima. On other hand, the function $J(\ee){=}{\mathrm{sinc}}(\ee)$ is characterized by infinite number of critical values $J(\ee')$ where $\ee'$ are all possible solutions of equation $\der{}{\ee}{\mathrm{sinc}(\ee)}{=}0$. Nevertheless, all these critical values still constitute a countable set which therefore is a null subset of all possible function's values. Needless to say, the function ${\mathrm{sinc}}(\ee)$ has infinite number of local minima an maxima.

Thus, satisfaction of theorem~\ref{__theo_PT2} does not rely on presence or absence of local minima and maxima in the constrained landscape $J(\eee|\ee_j{=}\cc)$ and, hence, cannot imply theorem~\ref{__theo_rsl_b} as a corollary. Let us illustrate this fact and explicitly disprove the theorem~\ref{__theo_rsl_b} by two simple counterexamples.

\begin{figure}[tbp]
\centering\includegraphics[width=0.65\columnwidth] {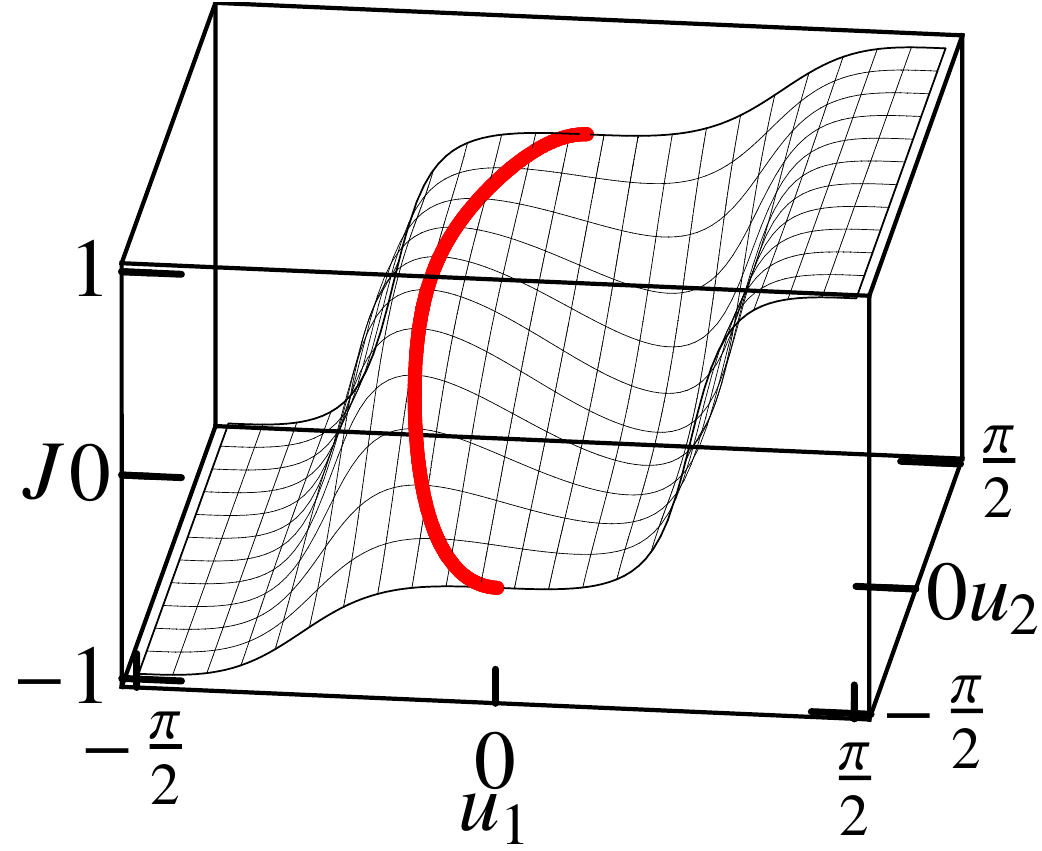}
\caption{The landscape $J(\ee_1,\ee_2)$ defined by equation~\eqref{__example_landscape}.
\label{@FIG.01}}
\end{figure}

The first illustrative counterexample is adopted from Ref.~\cite{2017-Zhdanov3}. Consider the trap-free landscape of two control parameters $\ee_1{\in}[-\frac{\pi}2,\frac{\pi}2]$ and $\ee_2{\in}[-\frac{\pi}2,\frac{\pi}2]$ defined by formula
\begin{gather}\label{__example_landscape}
J(\ee_1,\ee_2)=\frac2{\pi}\left(\tan(\ee_1)^3{-}\tan(\ee_1)\cos(\ee_2){+}\tan\left(\frac{\ee_2}{2}\right)\right).
\end{gather}
The landscape \eqref{__example_landscape} is shown in Fig.~\ref{@FIG.01}. 
It fulfills the local surjectivity condition of theorem \ref{__theo_rsl_b} and obeys parametric transversality theorem~\ref{__theo_PT2}. For instance, each value $J_0{\in}[-1,1]$ corresponds to at most two values $c$ such that $J_0$ is a local extremum of the constrained landscape $J(\ee_1|\ee_2{=}c)$. At the same time, the conclusion of theorem \ref{__theo_rsl_b} is violated: for \emph{any} $c{\in}(-\frac{\pi}2,\frac{\pi}2)$ the corresponding constrained landscape $J(\ee_1|\ee_2{=}c)$ has a local maximum. The set of these maxima for different values of $c$ is indicated by thick red curve in Fig.~\ref{@FIG.01}. 

Our second, more general counterexample is aimed to show that the conclusion of theorem \ref{__theo_rsl_b} doesn't follow from satisfaction of theorem \ref{__theo_PT2} for a broad class of functions. Specifically, consider arbitrary smooth locally surjective functions $\JJ(\eee)$ which acquire a finite number of local extrema after fixing one of the control parameters $\ee_{j}$ to any value from certain finite interval. Let us show that nearly all such functions satisfy theorem \ref{__theo_PT2}. Indeed, suppose that the function $J(\eee|\ee_{j}{=}c)$ has $S$ critical points which we will denote as $\eeecr_s$ ($s{=}0,...S{-}1$). For simplicity, let us assume that there are no saddle points among them. Without loss of generality, let us assume that $J(\eeecr_{s}){=}J_0$ for $0{\leq}s{\leq}s_0$, and $J(\eeecr_{s}){\ne}J_0$ for $s{>}s_0$. Denote $\Delta J_{0}{=}\min_{s{>}s_0}|J(\eeecr_{s}){-}J_0|$. Consider the family of constrained landscapes $J(\eee|\ee_{j}{=}c{+}\delta c)$ for some small variations of control parameters. Let us denote the corresponding critical points and their images as $\eeecr_{s}(\delta c)$ and $\JJcr_{s}(\delta c)$, respectively. Since $J(\eee)$ is smooth it is always possible to find such positive constant $\delta c_{\idx{iso}}$ that
\begin{enumerate}
\item $|\JJcr_{s}(\delta c){-}\JJcr_{s}(0)|{<}\frac{\Delta J_{0}}2$ for all $|\delta c|<\delta c_{\idx{iso}}$, and
\item all the landscapes in the family $|\delta c|<\delta c_{\idx{iso}}$ have the same number of critical points, and the maps $\delta c{\to}\eeecr_{s}(\delta c)$ and $\delta c{\to}\JJcr_{s}(\delta c)$ are smooth.
\end{enumerate}
Note that $\JJcr_{s}(\delta c)$ cannot be constant functions of $\delta c$ because the corresponding point $\eeecr_{s}$ would be a local extremum of $\JJ(\eee)$ which contradicts to our assumption. Hence, one can find some constant $\delta c_{\idx{max}}{\leq}\delta c_{\idx{iso}}$, such that $\JJcr_{s}(\delta c){\ne}J_0$ for all such $\delta c$ that $0{<}|\delta c|{<}\delta c_{\idx{max}}$. We conclude that if the statements 1) and 2) of theorem \ref{__theo_PT2} are satisfied for certain $J_0$ and $\ee_j{=}c$ then this is the only solution for $\ee_j$ on interval $(c{-}\delta c_{\idx{max}},c{+}\delta c_{\idx{max}})$. This means that all the critical values $J_0$ are isolated points, and hence constitute a null set. Thus, the assumptions and conclusions of theorem \ref{__theo_PT2} do not conflict with the existence of local extrema of $\JJ(\eee|\ee_j{=}c)$.

The case when some of the critical points of $\JJ(\eee|\ee_j{=}c)$ are saddle points can be treated similarly. However, an additional caution is required in this case because the variation $\delta c$ of constraint might change the number of critical points by annihilating a saddle point or via splitting it into new local extrema. Nevertheless, the net qualitative conclusion remains the same.

\section{Conclusion}
The conjecture that quantum control landscapes are almost always trap free remains the subject of ongoing debates since the initial attempt to prove it in 2004 \cite{2004-Rabitz} (see, e.g., the discussion \cite{2011-Pechen-Rabiz-reply,2012-Pechen} following the paper \cite{2011-Pechen}; a brief review also can be found in Ref.~\cite{2017-Zhdanov3}). In this comment we have shown that the new attempt to prove it made in Ref.~\cite{2017-Russell} also doesn't stand up to criticism. We identified two mistakes in the proposed proof. The origin of the first one is in overlooked effect of control boundedness. The second mistake stems from incorrect application of parametric transversality theorem. 

The authors of Ref.~\cite{2017-Russell} broadly applied the discussed results in their subsequent publications \cite{2017-Russell2,2018-Russell} to explain a variety of experimental evidences of trap-free landscapes in the areas far beyond the scope of quantum control, such as chemical synthesis, property optimization and evolutionary biology. It is worth stressing that these explanations are derived from the argument summarized as theorem~\ref{__theo_rsl_b} in this comment, which we have shown by counterexamples to be incorrect.

As a final remark, let us note that the statements of the authors of Ref.~\cite{2017-Russell} about ``mounting numerical evidence'' of ``common ease of quantum control optimization'' and ``the evident rarity of landscape traps'' are somewhat challenged by a number of results not mentioned in Ref.~\cite{2017-Russell}. Specifically, we would like to point out a series of works by D'Alessandro \cite{2015-Albertini,2016-Romano}, Boscain \cite{BOOK-Boscain,2002-Boscain,2005-Boscain,2006-Boscain,2014-Boscain}, Bonnard, Sugny (with supporting experiments by Glaser group) \cite{2012-Bonnard,2013-Garon,2014-Van_Damme,2017-Van_Damme,2017-Van_Reeth} and others \cite{2012-Salamon,2015-Zhdanov}. It is also worth mentioning a recent conjecture that proposition \ref{@P:key_result} is almost always violated for complex disordered quantum systems \cite{2017-Yang}.

\bibliography{RabitzGate_SCI_T}

\begin{thebibliography}{29}%
\makeatletter
\providecommand \@ifxundefined [1]{%
 \@ifx{#1\undefined}
}%
\providecommand \@ifnum [1]{%
 \ifnum #1\expandafter \@firstoftwo
 \else \expandafter \@secondoftwo
 \fi
}%
\providecommand \@ifx [1]{%
 \ifx #1\expandafter \@firstoftwo
 \else \expandafter \@secondoftwo
 \fi
}%
\providecommand \natexlab [1]{#1}%
\providecommand \enquote  [1]{``#1''}%
\providecommand \bibnamefont  [1]{#1}%
\providecommand \bibfnamefont [1]{#1}%
\providecommand \citenamefont [1]{#1}%
\providecommand \href@noop [0]{\@secondoftwo}%
\providecommand \href [0]{\begingroup \@sanitize@url \@href}%
\providecommand \@href[1]{\@@startlink{#1}\@@href}%
\providecommand \@@href[1]{\endgroup#1\@@endlink}%
\providecommand \@sanitize@url [0]{\catcode `\\12\catcode `\$12\catcode
  `\&12\catcode `\#12\catcode `\^12\catcode `\_12\catcode `\%12\relax}%
\providecommand \@@startlink[1]{}%
\providecommand \@@endlink[0]{}%
\providecommand \url  [0]{\begingroup\@sanitize@url \@url }%
\providecommand \@url [1]{\endgroup\@href {#1}{\urlprefix }}%
\providecommand \urlprefix  [0]{URL }%
\providecommand \Eprint [0]{\href }%
\providecommand \doibase [0]{http://dx.doi.org/}%
\providecommand \selectlanguage [0]{\@gobble}%
\providecommand \bibinfo  [0]{\@secondoftwo}%
\providecommand \bibfield  [0]{\@secondoftwo}%
\providecommand \translation [1]{[#1]}%
\providecommand \BibitemOpen [0]{}%
\providecommand \bibitemStop [0]{}%
\providecommand \bibitemNoStop [0]{.\EOS\space}%
\providecommand \EOS [0]{\spacefactor3000\relax}%
\providecommand \BibitemShut  [1]{\csname bibitem#1\endcsname}%
\let\auto@bib@innerbib\@empty
\bibitem [{\citenamefont {Russell}\ \emph
  {et~al.}(2018{\natexlab{a}})\citenamefont {Russell}, \citenamefont {Wu},\
  and\ \citenamefont {Rabitz}}]{2018-Russell-a}%
  \BibitemOpen
  \bibfield  {author} {\bibinfo {author} {\bibfnamefont {B.~J.}\ \bibnamefont
  {Russell}}, \bibinfo {author} {\bibfnamefont {R.-B.}\ \bibnamefont {Wu}}, \
  and\ \bibinfo {author} {\bibfnamefont {H.}~\bibnamefont {Rabitz}},\
  }\bibfield  {title} {\enquote {\bibinfo {title} {Reply to comment on "control
  landscapes are almost always trap free: a geometric assessment"},}\ }\href
  {\doibase 10.1088/1751-8121/aaecf2} {\bibfield  {journal} {\bibinfo
  {journal} {J. Phys. A: Math. Theor}\ } (\bibinfo {year}
  {2018}{\natexlab{a}}),\ 10.1088/1751-8121/aaecf2}\BibitemShut {NoStop}%
\bibitem [{\citenamefont {Russell}\ \emph {et~al.}(2017)\citenamefont
  {Russell}, \citenamefont {Rabitz},\ and\ \citenamefont {Wu}}]{2017-Russell}%
  \BibitemOpen
  \bibfield  {author} {\bibinfo {author} {\bibfnamefont {B.}~\bibnamefont
  {Russell}}, \bibinfo {author} {\bibfnamefont {H.}~\bibnamefont {Rabitz}}, \
  and\ \bibinfo {author} {\bibfnamefont {R.-B.}\ \bibnamefont {Wu}},\
  }\bibfield  {title} {\enquote {\bibinfo {title} {Control landscapes are
  almost always trap free: a geometric assessment},}\ }\href {\doibase
  10.1088/1751-8121/aa6b77} {\bibfield  {journal} {\bibinfo  {journal} {J.
  Phys. A: Math. Theor.}\ }\textbf {\bibinfo {volume} {50}},\ \bibinfo {pages}
  {205302} (\bibinfo {year} {2017})}\BibitemShut {NoStop}%
\bibitem [{\citenamefont {Kosut}\ \emph {et~al.}(2018)\citenamefont {Kosut},
  \citenamefont {Arenz},\ and\ \citenamefont {Rabitz}}]{2018-Kosut}%
  \BibitemOpen
  \bibfield  {author} {\bibinfo {author} {\bibfnamefont {R.~L.}\ \bibnamefont
  {Kosut}}, \bibinfo {author} {\bibfnamefont {C.}~\bibnamefont {Arenz}}, \ and\
  \bibinfo {author} {\bibfnamefont {H.}~\bibnamefont {Rabitz}},\ }\bibfield
  {title} {\enquote {\bibinfo {title} {Quantum control landscape of bipartite
  systems},}\ }\href {https://arxiv.org/abs/1810.04362} {\bibfield  {journal}
  {\bibinfo  {journal} {arXiv:1810.04362 [quant-ph]}\ } (\bibinfo {year}
  {2018})}\BibitemShut {NoStop}%
\bibitem [{\citenamefont {Vakulenko}(2014)}]{2014-Vakulenko}%
  \BibitemOpen
  \bibfield  {author} {\bibinfo {author} {\bibfnamefont {S.}~\bibnamefont
  {Vakulenko}},\ }\href@noop {} {\emph {\bibinfo {title} {{Complexity and
  Evolution of Dissipative Systems: An Analytical Approach}}}},\ \bibinfo
  {series} {{De Gruyter Series in Mathematics and Life Sciences}},
  Vol.~\bibinfo {volume} {4}\ (\bibinfo  {publisher} {Walter de Gruyter},\
  \bibinfo {year} {2014})\BibitemShut {NoStop}%
\bibitem [{\citenamefont {von Neumann}(1937)}]{1937-von_Neumann}%
  \BibitemOpen
  \bibfield  {author} {\bibinfo {author} {\bibfnamefont {J.}~\bibnamefont {von
  Neumann}},\ }\bibfield  {title} {\enquote {\bibinfo {title} {Some matrix
  inequalities and metrization of metric space},}\ }\href@noop {} {\bibfield
  {journal} {\bibinfo  {journal} {Tomsk Univ. Rev}\ }\textbf {\bibinfo {volume}
  {1}},\ \bibinfo {pages} {286--296} (\bibinfo {year} {1937})}\BibitemShut
  {NoStop}%
\bibitem [{\citenamefont {Brockett}(1988)}]{1988-Brockett}%
  \BibitemOpen
  \bibfield  {author} {\bibinfo {author} {\bibfnamefont {R.~W.}\ \bibnamefont
  {Brockett}},\ }\bibfield  {title} {\enquote {\bibinfo {title} {Dynamical
  systems that sort lists, diagonalize matrices and solve linear programming
  problems},}\ }in\ \href {\doibase 10.1109/CDC.1988.194420} {\emph {\bibinfo
  {booktitle} {Proceedings of the 27th IEEE Conference on Decision and
  Control}}},\ Vol.~\bibinfo {volume} {1}\ (\bibinfo {year} {1988})\ pp.\
  \bibinfo {pages} {799--803}\BibitemShut {NoStop}%
\bibitem [{\citenamefont {Russell}\ and\ \citenamefont
  {Rabitz}(2017)}]{2017-Russell2}%
  \BibitemOpen
  \bibfield  {author} {\bibinfo {author} {\bibfnamefont {B.}~\bibnamefont
  {Russell}}\ and\ \bibinfo {author} {\bibfnamefont {H.}~\bibnamefont
  {Rabitz}},\ }\bibfield  {title} {\enquote {\bibinfo {title} {Common
  foundations of optimal control across the sciences: evidence of a free
  lunch},}\ }\href {\doibase 10.1098/rsta.2016.0210} {\bibfield  {journal}
  {\bibinfo  {journal} {Phil. Trans. R. Soc. A}\ }\textbf {\bibinfo {volume}
  {375}},\ \bibinfo {pages} {20160210} (\bibinfo {year} {2017})}\BibitemShut
  {NoStop}%
\bibitem [{\citenamefont {Russell}\ \emph
  {et~al.}(2018{\natexlab{b}})\citenamefont {Russell}, \citenamefont {Vuglar},\
  and\ \citenamefont {Rabitz}}]{2018-Russell}%
  \BibitemOpen
  \bibfield  {author} {\bibinfo {author} {\bibfnamefont {B.}~\bibnamefont
  {Russell}}, \bibinfo {author} {\bibfnamefont {S.}~\bibnamefont {Vuglar}}, \
  and\ \bibinfo {author} {\bibfnamefont {H.}~\bibnamefont {Rabitz}},\
  }\bibfield  {title} {\enquote {\bibinfo {title} {Control landscapes for a
  class of non-linear dynamical systems: sufficient conditions for the absence
  of traps},}\ }\href {\doibase 10.1088/1751-8121/aacc85} {\bibfield  {journal}
  {\bibinfo  {journal} {J. Phys. A: Math. Theor.}\ }\textbf {\bibinfo {volume}
  {51}},\ \bibinfo {pages} {335103} (\bibinfo {year}
  {2018}{\natexlab{b}})}\BibitemShut {NoStop}%
\bibitem [{\citenamefont {Arnol'd}\ \emph {et~al.}(2012)\citenamefont
  {Arnol'd}, \citenamefont {Gusein-Zade},\ and\ \citenamefont
  {Varchenko}}]{1985-Arnol'd}%
  \BibitemOpen
  \bibfield  {author} {\bibinfo {author} {\bibfnamefont {V.~I.}\ \bibnamefont
  {Arnol'd}}, \bibinfo {author} {\bibfnamefont {S.~M.}\ \bibnamefont
  {Gusein-Zade}}, \ and\ \bibinfo {author} {\bibfnamefont {A.~N.}\ \bibnamefont
  {Varchenko}},\ }\href@noop {} {\emph {\bibinfo {title} {{Singularities of
  differentiable maps. Volume 1. Classification of Critical Points, Caustics
  and Wave Fronts}}}},\ {Modern Birkh\"{a}user Classics}\ (\bibinfo
  {publisher} {Birkh\"{a}user},\ \bibinfo {address} {New York},\ \bibinfo
  {year} {2012})\BibitemShut {NoStop}%
\bibitem [{\citenamefont {Zhdanov}(2017)}]{2017-Zhdanov3}%
  \BibitemOpen
  \bibfield  {author} {\bibinfo {author} {\bibfnamefont {D.~V.}\ \bibnamefont
  {Zhdanov}},\ }\bibfield  {title} {\enquote {\bibinfo {title} {Theory of
  quantum control landscapes: Overlooked hidden cracks},}\ }\href
  {http://arxiv.org/abs/1710.07753} {\bibfield  {journal} {\bibinfo  {journal}
  {arXiv:1710.07753 [quant-ph]}\ } (\bibinfo {year} {2017})}\BibitemShut
  {NoStop}%
\bibitem [{\citenamefont {Rabitz}\ \emph {et~al.}(2004)\citenamefont {Rabitz},
  \citenamefont {Hsieh},\ and\ \citenamefont {Rosenthal}}]{2004-Rabitz}%
  \BibitemOpen
  \bibfield  {author} {\bibinfo {author} {\bibfnamefont {H.~A.}\ \bibnamefont
  {Rabitz}}, \bibinfo {author} {\bibfnamefont {M.~M.}\ \bibnamefont {Hsieh}}, \
  and\ \bibinfo {author} {\bibfnamefont {C.~M.}\ \bibnamefont {Rosenthal}},\
  }\bibfield  {title} {\enquote {\bibinfo {title} {Quantum optimally controlled
  transition landscapes},}\ }\href {\doibase 10.1126/science.1093649}
  {\bibfield  {journal} {\bibinfo  {journal} {Science}\ }\textbf {\bibinfo
  {volume} {303}},\ \bibinfo {pages} {1998--2001} (\bibinfo {year}
  {2004})}\BibitemShut {NoStop}%
\bibitem [{\citenamefont {Rabitz}\ \emph {et~al.}(2012)\citenamefont {Rabitz},
  \citenamefont {Ho}, \citenamefont {Long}, \citenamefont {Wu},\ and\
  \citenamefont {Brif}}]{2011-Pechen-Rabiz-reply}%
  \BibitemOpen
  \bibfield  {author} {\bibinfo {author} {\bibfnamefont {H.}~\bibnamefont
  {Rabitz}}, \bibinfo {author} {\bibfnamefont {T.-S.}\ \bibnamefont {Ho}},
  \bibinfo {author} {\bibfnamefont {R.}~\bibnamefont {Long}}, \bibinfo {author}
  {\bibfnamefont {R.}~\bibnamefont {Wu}}, \ and\ \bibinfo {author}
  {\bibfnamefont {C.}~\bibnamefont {Brif}},\ }\bibfield  {title} {\enquote
  {\bibinfo {title} {Comment on ``are there traps in quantum control
  landscapes?''},}\ }\href {\doibase 10.1103/PhysRevLett.108.198901} {\bibfield
   {journal} {\bibinfo  {journal} {Phys. Rev. Lett.}\ }\textbf {\bibinfo
  {volume} {108}},\ \bibinfo {pages} {198901} (\bibinfo {year}
  {2012})}\BibitemShut {NoStop}%
\bibitem [{\citenamefont {Pechen}\ and\ \citenamefont
  {Tannor}(2012)}]{2012-Pechen}%
  \BibitemOpen
  \bibfield  {author} {\bibinfo {author} {\bibfnamefont {A.~N.}\ \bibnamefont
  {Pechen}}\ and\ \bibinfo {author} {\bibfnamefont {D.~J.}\ \bibnamefont
  {Tannor}},\ }\bibfield  {title} {\enquote {\bibinfo {title} {Pechen and
  tannor reply:},}\ }\href {\doibase 10.1103/PhysRevLett.108.198902} {\bibfield
   {journal} {\bibinfo  {journal} {Phys. Rev. Lett.}\ }\textbf {\bibinfo
  {volume} {108}},\ \bibinfo {pages} {198902} (\bibinfo {year}
  {2012})}\BibitemShut {NoStop}%
\bibitem [{\citenamefont {Pechen}\ and\ \citenamefont
  {Tannor}(2011)}]{2011-Pechen}%
  \BibitemOpen
  \bibfield  {author} {\bibinfo {author} {\bibfnamefont {A.}~\bibnamefont
  {Pechen}}\ and\ \bibinfo {author} {\bibfnamefont {D.}~\bibnamefont
  {Tannor}},\ }\bibfield  {title} {\enquote {\bibinfo {title} {Are there traps
  in quantum control landscapes?}}\ }\href {\doibase
  10.1103/PhysRevLett.106.120402} {\bibfield  {journal} {\bibinfo  {journal}
  {Phys. Rev. Lett.}\ }\textbf {\bibinfo {volume} {106}},\ \bibinfo {pages}
  {120402} (\bibinfo {year} {2011})}\BibitemShut {NoStop}%
\bibitem [{\citenamefont {Albertini}\ and\ \citenamefont
  {D'Alessandro}(2015)}]{2015-Albertini}%
  \BibitemOpen
  \bibfield  {author} {\bibinfo {author} {\bibfnamefont {F.}~\bibnamefont
  {Albertini}}\ and\ \bibinfo {author} {\bibfnamefont {D.}~\bibnamefont
  {D'Alessandro}},\ }\bibfield  {title} {\enquote {\bibinfo {title} {Minimum
  time optimal synthesis for two level quantum systems},}\ }\href {\doibase
  10.1063/1.4906137} {\bibfield  {journal} {\bibinfo  {journal} {J. Math.
  Phys.}\ }\textbf {\bibinfo {volume} {56}},\ \bibinfo {pages} {012106}
  (\bibinfo {year} {2015})}\BibitemShut {NoStop}%
\bibitem [{\citenamefont {Romano}\ and\ \citenamefont
  {D'Alessandro}(2016)}]{2016-Romano}%
  \BibitemOpen
  \bibfield  {author} {\bibinfo {author} {\bibfnamefont {R.}~\bibnamefont
  {Romano}}\ and\ \bibinfo {author} {\bibfnamefont {D.}~\bibnamefont
  {D'Alessandro}},\ }\bibfield  {title} {\enquote {\bibinfo {title} {Minimum
  time control of a pair of two-level quantum systems with opposite drifts},}\
  }\href {\doibase 10.1088/1751-8113/49/34/345303} {\bibfield  {journal}
  {\bibinfo  {journal} {J. Phys. A: Math. Theor.}\ }\textbf {\bibinfo {volume}
  {49}},\ \bibinfo {pages} {345303} (\bibinfo {year} {2016})}\BibitemShut
  {NoStop}%
\bibitem [{\citenamefont {Boscain}\ and\ \citenamefont
  {Piccoli}(2003)}]{BOOK-Boscain}%
  \BibitemOpen
  \bibfield  {author} {\bibinfo {author} {\bibfnamefont {U.}~\bibnamefont
  {Boscain}}\ and\ \bibinfo {author} {\bibfnamefont {B.}~\bibnamefont
  {Piccoli}},\ }\href@noop {} {\emph {\bibinfo {title} {{Optimal syntheses for
  control systems on 2-D manifolds}}}},\ \bibinfo {series} {{Mathematiques et
  Applications}}, Vol.~\bibinfo {volume} {43}\ (\bibinfo  {publisher} {Springer
  Science \& Business Media},\ \bibinfo {year} {2003})\BibitemShut {NoStop}%
\bibitem [{\citenamefont {Boscain}\ \emph {et~al.}(2002)\citenamefont
  {Boscain}, \citenamefont {Charlot}, \citenamefont {Gauthier}, \citenamefont
  {Guerin},\ and\ \citenamefont {Jauslin}}]{2002-Boscain}%
  \BibitemOpen
  \bibfield  {author} {\bibinfo {author} {\bibfnamefont {U.}~\bibnamefont
  {Boscain}}, \bibinfo {author} {\bibfnamefont {G.}~\bibnamefont {Charlot}},
  \bibinfo {author} {\bibfnamefont {J.-P.}\ \bibnamefont {Gauthier}}, \bibinfo
  {author} {\bibfnamefont {S.}~\bibnamefont {Guerin}}, \ and\ \bibinfo {author}
  {\bibfnamefont {H.-R.}\ \bibnamefont {Jauslin}},\ }\bibfield  {title}
  {\enquote {\bibinfo {title} {Optimal control in laser-induced population
  transfer for two- and three-level quantum systems},}\ }\href {\doibase
  10.1063/1.1465516} {\bibfield  {journal} {\bibinfo  {journal} {J. Math.
  Phys.}\ }\textbf {\bibinfo {volume} {43}},\ \bibinfo {pages} {2107} (\bibinfo
  {year} {2002})}\BibitemShut {NoStop}%
\bibitem [{\citenamefont {Boscain}\ and\ \citenamefont
  {Chitour}(2005)}]{2005-Boscain}%
  \BibitemOpen
  \bibfield  {author} {\bibinfo {author} {\bibfnamefont {U.}~\bibnamefont
  {Boscain}}\ and\ \bibinfo {author} {\bibfnamefont {Y.}~\bibnamefont
  {Chitour}},\ }\bibfield  {title} {\enquote {\bibinfo {title} {{Time-optimal
  synthesis for left-invariant control systems on SO(3)}},}\ }\href {\doibase
  10.1137/S0363012904441532} {\bibfield  {journal} {\bibinfo  {journal} {SIAM
  J. Control}\ }\textbf {\bibinfo {volume} {44}},\ \bibinfo {pages} {111--139}
  (\bibinfo {year} {2005})}\BibitemShut {NoStop}%
\bibitem [{\citenamefont {Boscain}\ and\ \citenamefont
  {Mason}(2006)}]{2006-Boscain}%
  \BibitemOpen
  \bibfield  {author} {\bibinfo {author} {\bibfnamefont {U.}~\bibnamefont
  {Boscain}}\ and\ \bibinfo {author} {\bibfnamefont {P.}~\bibnamefont
  {Mason}},\ }\bibfield  {title} {\enquote {\bibinfo {title} {Time minimal
  trajectories for a spin 1/2 particle in a magnetic field},}\ }\href {\doibase
  10.1063/1.2203236} {\bibfield  {journal} {\bibinfo  {journal} {J. Math.
  Phys.}\ }\textbf {\bibinfo {volume} {47}},\ \bibinfo {pages} {062101}
  (\bibinfo {year} {2006})}\BibitemShut {NoStop}%
\bibitem [{\citenamefont {Boscain}\ \emph {et~al.}(2014)\citenamefont
  {Boscain}, \citenamefont {Gronberg}, \citenamefont {Long},\ and\
  \citenamefont {Rabitz}}]{2014-Boscain}%
  \BibitemOpen
  \bibfield  {author} {\bibinfo {author} {\bibfnamefont {U.}~\bibnamefont
  {Boscain}}, \bibinfo {author} {\bibfnamefont {F.}~\bibnamefont {Gronberg}},
  \bibinfo {author} {\bibfnamefont {R.}~\bibnamefont {Long}}, \ and\ \bibinfo
  {author} {\bibfnamefont {H.}~\bibnamefont {Rabitz}},\ }\bibfield  {title}
  {\enquote {\bibinfo {title} {Minimal time trajectories for two-level quantum
  systems with two bounded controls},}\ }\href {\doibase 10.1063/1.4882158}
  {\bibfield  {journal} {\bibinfo  {journal} {J. Math. Phys.}\ }\textbf
  {\bibinfo {volume} {55}},\ \bibinfo {pages} {062106} (\bibinfo {year}
  {2014})}\BibitemShut {NoStop}%
\bibitem [{\citenamefont {Bonnard}\ \emph {et~al.}(2012)\citenamefont
  {Bonnard}, \citenamefont {Glaser},\ and\ \citenamefont
  {Sugny}}]{2012-Bonnard}%
  \BibitemOpen
  \bibfield  {author} {\bibinfo {author} {\bibfnamefont {B.}~\bibnamefont
  {Bonnard}}, \bibinfo {author} {\bibfnamefont {S.~J.}\ \bibnamefont {Glaser}},
  \ and\ \bibinfo {author} {\bibfnamefont {D.}~\bibnamefont {Sugny}},\
  }\bibfield  {title} {\enquote {\bibinfo {title} {A review of geometric
  optimal control for quantum systems in nuclear magnetic resonance},}\ }\href
  {\doibase 10.1155/2012/857493} {\bibfield  {journal} {\bibinfo  {journal}
  {Adv. Math. Phys.}\ }\textbf {\bibinfo {volume} {2012}},\ \bibinfo {pages}
  {1--29} (\bibinfo {year} {2012})}\BibitemShut {NoStop}%
\bibitem [{\citenamefont {Garon}\ \emph {et~al.}(2013)\citenamefont {Garon},
  \citenamefont {Glaser},\ and\ \citenamefont {Sugny}}]{2013-Garon}%
  \BibitemOpen
  \bibfield  {author} {\bibinfo {author} {\bibfnamefont {A.}~\bibnamefont
  {Garon}}, \bibinfo {author} {\bibfnamefont {S.~J.}\ \bibnamefont {Glaser}}, \
  and\ \bibinfo {author} {\bibfnamefont {D.}~\bibnamefont {Sugny}},\ }\bibfield
   {title} {\enquote {\bibinfo {title} {{Time-optimal control of SU(2) quantum
  operations}},}\ }\href {\doibase 10.1103/PhysRevA.88.043422} {\bibfield
  {journal} {\bibinfo  {journal} {Phys. Rev. A}\ }\textbf {\bibinfo {volume}
  {88}},\ \bibinfo {pages} {043422} (\bibinfo {year} {2013})}\BibitemShut
  {NoStop}%
\bibitem [{\citenamefont {Van~Damme}\ \emph {et~al.}(2014)\citenamefont
  {Van~Damme}, \citenamefont {Zeier}, \citenamefont {Glaser},\ and\
  \citenamefont {Sugny}}]{2014-Van_Damme}%
  \BibitemOpen
  \bibfield  {author} {\bibinfo {author} {\bibfnamefont {L.}~\bibnamefont
  {Van~Damme}}, \bibinfo {author} {\bibfnamefont {R.}~\bibnamefont {Zeier}},
  \bibinfo {author} {\bibfnamefont {S.~J.}\ \bibnamefont {Glaser}}, \ and\
  \bibinfo {author} {\bibfnamefont {D.}~\bibnamefont {Sugny}},\ }\bibfield
  {title} {\enquote {\bibinfo {title} {Application of the pontryagin maximum
  principle to the time-optimal control in a chain of three spins with unequal
  couplings},}\ }\href {\doibase 10.1103/PhysRevA.90.013409} {\bibfield
  {journal} {\bibinfo  {journal} {Phys. Rev. A}\ }\textbf {\bibinfo {volume}
  {90}},\ \bibinfo {pages} {013409} (\bibinfo {year} {2014})}\BibitemShut
  {NoStop}%
\bibitem [{\citenamefont {Van~Damme}\ \emph {et~al.}(2017)\citenamefont
  {Van~Damme}, \citenamefont {Ansel}, \citenamefont {Glaser},\ and\
  \citenamefont {Sugny}}]{2017-Van_Damme}%
  \BibitemOpen
  \bibfield  {author} {\bibinfo {author} {\bibfnamefont {L.}~\bibnamefont
  {Van~Damme}}, \bibinfo {author} {\bibfnamefont {Q.}~\bibnamefont {Ansel}},
  \bibinfo {author} {\bibfnamefont {S.~J.}\ \bibnamefont {Glaser}}, \ and\
  \bibinfo {author} {\bibfnamefont {D.}~\bibnamefont {Sugny}},\ }\bibfield
  {title} {\enquote {\bibinfo {title} {Robust optimal control of two-level
  quantum systems},}\ }\href {\doibase 10.1103/PhysRevA.95.063403} {\bibfield
  {journal} {\bibinfo  {journal} {Phys. Rev. A}\ }\textbf {\bibinfo {volume}
  {95}},\ \bibinfo {pages} {063403} (\bibinfo {year} {2017})}\BibitemShut
  {NoStop}%
\bibitem [{\citenamefont {Van~Reeth}\ \emph {et~al.}(2017)\citenamefont
  {Van~Reeth}, \citenamefont {Ratiney}, \citenamefont {Tesch}, \citenamefont
  {Grenier}, \citenamefont {Beuf}, \citenamefont {Glaser},\ and\ \citenamefont
  {Sugny}}]{2017-Van_Reeth}%
  \BibitemOpen
  \bibfield  {author} {\bibinfo {author} {\bibfnamefont {E.}~\bibnamefont
  {Van~Reeth}}, \bibinfo {author} {\bibfnamefont {H.}~\bibnamefont {Ratiney}},
  \bibinfo {author} {\bibfnamefont {M.}~\bibnamefont {Tesch}}, \bibinfo
  {author} {\bibfnamefont {D.}~\bibnamefont {Grenier}}, \bibinfo {author}
  {\bibfnamefont {O.}~\bibnamefont {Beuf}}, \bibinfo {author} {\bibfnamefont
  {S.~J.}\ \bibnamefont {Glaser}}, \ and\ \bibinfo {author} {\bibfnamefont
  {D.}~\bibnamefont {Sugny}},\ }\bibfield  {title} {\enquote {\bibinfo {title}
  {{Optimal control design of preparation pulses for contrast optimization in
  MRI}},}\ }\href {\doibase 10.1016/j.jmr.2017.04.012} {\bibfield  {journal}
  {\bibinfo  {journal} {J. Magn. Reson.}\ }\textbf {\bibinfo {volume} {279}},\
  \bibinfo {pages} {39--50} (\bibinfo {year} {2017})}\BibitemShut {NoStop}%
\bibitem [{\citenamefont {Salamon}\ \emph {et~al.}(2012)\citenamefont
  {Salamon}, \citenamefont {Hoffmann},\ and\ \citenamefont
  {Tsirlin}}]{2012-Salamon}%
  \BibitemOpen
  \bibfield  {author} {\bibinfo {author} {\bibfnamefont {P.}~\bibnamefont
  {Salamon}}, \bibinfo {author} {\bibfnamefont {K.~H.}\ \bibnamefont
  {Hoffmann}}, \ and\ \bibinfo {author} {\bibfnamefont {A.}~\bibnamefont
  {Tsirlin}},\ }\bibfield  {title} {\enquote {\bibinfo {title} {Optimal control
  in a quantum cooling problem},}\ }\href {\doibase 10.1016/j.aml.2011.11.020}
  {\bibfield  {journal} {\bibinfo  {journal} {Appl. Math. Lett.}\ }\textbf
  {\bibinfo {volume} {25}},\ \bibinfo {pages} {1263--1266} (\bibinfo {year}
  {2012})}\BibitemShut {NoStop}%
\bibitem [{\citenamefont {Zhdanov}\ and\ \citenamefont
  {Seideman}(2015)}]{2015-Zhdanov}%
  \BibitemOpen
  \bibfield  {author} {\bibinfo {author} {\bibfnamefont {D.~V.}\ \bibnamefont
  {Zhdanov}}\ and\ \bibinfo {author} {\bibfnamefont {T.}~\bibnamefont
  {Seideman}},\ }\bibfield  {title} {\enquote {\bibinfo {title} {Role of
  control constraints in quantum optimal control},}\ }\href {\doibase
  10.1103/PhysRevA.92.052109} {\bibfield  {journal} {\bibinfo  {journal} {Phys.
  Rev. A}\ }\textbf {\bibinfo {volume} {92}},\ \bibinfo {pages} {052109}
  (\bibinfo {year} {2015})}\BibitemShut {NoStop}%
\bibitem [{\citenamefont {Yang}\ \emph {et~al.}(2017)\citenamefont {Yang},
  \citenamefont {Rahmani}, \citenamefont {Shabani}, \citenamefont {Neven},\
  and\ \citenamefont {Chamon}}]{2017-Yang}%
  \BibitemOpen
  \bibfield  {author} {\bibinfo {author} {\bibfnamefont {Z.-C.}\ \bibnamefont
  {Yang}}, \bibinfo {author} {\bibfnamefont {A.}~\bibnamefont {Rahmani}},
  \bibinfo {author} {\bibfnamefont {A.}~\bibnamefont {Shabani}}, \bibinfo
  {author} {\bibfnamefont {H.}~\bibnamefont {Neven}}, \ and\ \bibinfo {author}
  {\bibfnamefont {C.}~\bibnamefont {Chamon}},\ }\bibfield  {title} {\enquote
  {\bibinfo {title} {Optimizing variational quantum algorithms using
  pontryagin's minimum principle},}\ }\href {\doibase
  10.1103/PhysRevX.7.021027} {\bibfield  {journal} {\bibinfo  {journal} {Phys.
  Rev. X}\ }\textbf {\bibinfo {volume} {7}},\ \bibinfo {pages} {021027}
  (\bibinfo {year} {2017})}\BibitemShut {NoStop}%
\end{thebibliography}%

\end{document}